\documentstyle[preprint,aps]{revtex}
\begin{document}
\title{Charge, Orbital and Magnetic Order in $Nd_{0.5}Ca_{0.5}MnO_{3}$}
\author{F. Millange$^{1}$,S.\ de Brion$^{2\ast }$, G. Chouteau$^{2}$}
\address{$^{1}$ CRISMAT-ISMRA\\
CNRS\ and Universit\'{e} de Caen\\
6, boulevard du Mar\'{e}chal Juin\\
14050 Caen - France\\
present address: Inorganic Chemistry Laboratory \\
South Parks - Oxford OX1 3QR, England. \\
$^{2}$Grenoble High Magnetic Field Laboratory\\
BP 166 - 38042 Grenoble cedex 9- France\\
$^{\ast }$ corresponding author}
\maketitle
\date{\today }

\begin{abstract}
In the manganite $Nd_{0.5}Ca_{0.5}MnO_{3}$, charge ordering occurs at much
higher temperature than the antiferromagnetic order $(T_{CO}=250K,%
\,T_{N}=160K)$.\ the magnetic behavior of the phase $T_{N}<T<T_{CO}$ is
puzzling: its magnetization and susceptibility are typical of an
antiferromagnet while no magnetic order is detected by neutron
diffraction.We have undertaken an extensive study of the cristallographic,
electric and magnetic properties of $Nd_{0.5}Ca_{0.5}MnO_{3}$ and
established its phase diagram as a function of temperature and magnetic
field. The charge disordered, paramagnetic phase above $T_{CO\text{ }}$%
present ferromagnetic correlations.\ An antiferromagnetic CE phase prevails
below $T_{N}$, with complete charge and orbital ordering.\ In the
intermediate temperature range, charge ordering occurs while orbital
ordering sets in progressively, with no magnetic order.\ Strong magnetic
fields destroy the charge ordered phases in a fisrt order transition towards
a ferromagnetic state.
\end{abstract}

.

\newpage

\section{Introduction}

The manganites $L_{1-x}M_{x}MnO_{3}$, where $L$ is a trivalent rare earth
and $M$ a divalent alkaline earth element, present astonishing properties
due to the interplay of magnetism, electric transport and crystallographic
distortion. Much interest has been focused on the hole doped ferromagnetic
phase $(x\simeq 0.3)$ which exhibits colossal magnetoresistance\cite{Jin}.\
At fractional doping level, another mechanism is superimposed: charge
ordering between $Mn^{3+}$ and $Mn^{4+}$ ions takes place\cite{PrSr}\cite
{NdSr}\cite{Rao}\cite{caen1}.\ The crystallographic and magnetic environment
is then strongly modified. For instance, $Nd_{0.5}Sr_{0.5}MnO_{3}$ \cite
{NdSr}\cite{caen1} is ferromagnetic and metallic below 245K.\ At 160K, the
charge order transition occurs and the system becomes antiferromagnetic and
insulating. Applying a strong magnetic field restores the ferromagnetic
order together with the metallic conductivity.

However these properties depend crucially on the local lattice distortion
and the mismatch between the rare earth and the alkaline element\cite{Hwang} 
\cite{NdSm}. In these perovskite structure, it is convenient to define the
tolerance factor $t=\left( r_{L,M}+r_{O}\right) /\sqrt{2}\left(
r_{Mn}+r_{O}\right) $ where $r_{L,M}$, $r_{Mn}$ and $r_{O}$ are the average
ionic radii of the trivalent rare earth and divalent alkaline earth site,
manganese site and oxygen site respectively.\ It describes the distortion of
the perovskite structure. This distortion modifies the strength of the
different magnetic interactions present in these systems. The double
exchange mechanism\cite{double exchange} couples the itinerant $%
3d\,\,\,e_{g} $ electrons of $Mn^{3+}$ with the localized $3d\,t_{2g}$ spins
of the $Mn^{4+}$ ions.\ It favors ferromagnetism together with metallic
conductivity. It depends on the number of $e_{g}$ electrons as well as the
bond angle Mn-O-Mn governed by the lattice distortion. On\ the other hand,
the superexchange interaction between manganese ions favors
antiferromagnetic order with localized spins with then insulating electric
properties. Decreasing the ionic radius of the L and M site modifies the
bond angle Mn-O-Mn . The ferromagnetic ordering interaction is then reduced 
\cite{Hwang}.\ The bandwidth of the $e_{g}$ electron becomes narrower.
Eventually, the system becomes insulating and lost its ferromagnetic order.\ 

Additionally, for $x=0.5$, charge ordering will tend to localize the spins
and destabilize the ferromagnetic order. This is what has been observed in $%
Nd_{0.5}Sr_{0.5}MnO_{3}$\cite{NdSr} and in $Pr_{0.5}Sr_{0.5}MnO_{3}$\cite
{PrSr} where the tolerance factor is close to 0.95. The ferromagnetic,
metallic order, established below $T_{C}\approx 250K$, is destroyed\ when
charge ordering sets in at $T_{CO}\approx 150K$, and the system becomes
antiferromagnetic and insulating ($T_{N}=T_{CO})$.\ 

For smaller tolerance factor, the magnetic phase diagram is not so clear. In 
$Pr_{0.5}Ca_{0.5}MnO_{3}$ \cite{Pollert},\cite{PrCa}, with $t=0.933$,or in $%
Sm_{0.5}Ca_{0.5}MnO_{3}$ $(t=0.924)$\cite{Rao}, the charge order state
occurs at higher temperature ($T_{CO}\approx 250K$) so that no ferromagnetic
order is established in zero magnetic field.\ Antiferromagnetism is observed
only at lower temperature ($T_{N}\approx 150K<T_{CO})$.\ 

Surprisingly, the magnetic susceptibility presents a huge peak at $T_{CO}$,
as observed for a long range antiferromagnetic order, and none at $T_{N}$.
Such a behavior is very sensitive to stoichiometry. In $%
Pr_{0.6}Ca_{0.4}MnO_{3}$ \cite{Pollert},\cite{Okimoto}, as well as in $%
Nd_{0.6}Ca_{0.4}MnO_{3}$\cite{thesis}, the anomaly at $T_{CO}$ persists
while the signature of $T_{N}$ is quite visible.\ What is the exact nature
of the phase when $T_{N}<T<T_{CO}$? What is the mechanism responsible for
the susceptibility anomaly at $T_{CO}$? To answer these questions, we have
focused on the compound $Nd_{0.5}Ca_{0.5}MnO_{3}$ for which $t=0.930$.\ We
have undertaken an extensive study of its crystallographic, electric and
magnetic properties using neutron diffraction, electric transport, magnetic
susceptibility and high field magnetization measurements.\ 

\section{Experiments}

The powdered sample was prepared by solid state reaction from the component
oxides in appropriate ratios.\ The reagents were intimately mixed in an
agate mortar and heated in air at $900{{}^{\circ }}C$ for 12 hours to allow
decarbonation to take place. This powder was fired at $1200{{}^{\circ }}C$
for 12 hours. Final firing was performed at $1500{{}^{\circ }}C$ for five
days with intermediate grinding to ensure homogeneity. \ Phase purity was
checked by X ray diffraction using a Phillips diffractometer with Cu K$%
_{\alpha }$ radiation.

The phase $Nd_{0.5}Ca_{0.5}MnO_{3}$ was subsequently registered on the
high-resolution diffractometer D2b at the Institut Laue Langevin using a
wavelength of $\lambda =1.5938\,$\AA\ at room temperature and 1.5 K. The
sample was contained in a vanadium can. The high-flux powder diffractometer
CRG-CNRS D1b was used to investigate the thermal dependence of the magnetic
structure in the temperature range $1.5K-293K$. Thanks to the position
sensitive detector covering $80%
%TCIMACRO{\UNICODE[m]{0xb0}}%
%BeginExpansion
{{}^\circ}%
%EndExpansion
(2\theta )$ and the high neutron flux available at the $2.52\,$\AA\
wavelength, a neutron diffraction pattern has been measured every 6 Kelvin.
The nuclear and magnetic refinement were performed using the profile fitting
program Fullprof \cite{Fullprof} ($b_{Mn}=-0.373.10^{-12}cm$, $%
b_{O}=0.580.10^{-12}cm$, $b_{Nd}=0.769.10^{-12}cm$, $b_{Ca}=0.470.10^{-12}cm$%
).

Magnetic susceptibility measurements were made with a SQUID magnetometer
(MPMS Quantum Design) in the temperature range 4K-350K. These D.C.
susceptibility measurements were carried out at low field (10 mT) with
increasing temperature after the samples were either zero field cooled (ZFC)
or field cooled (FC). High temperature susceptibility measurements were also
taken up to 800K using a faraday balance. High field magnetization
measurements were performed, up to 22T, with a conventional extraction
set-up, in the temperature range 4K-300K.

The electric resistivity was measured between 5 K and 600 K, , using a
standard four points method, on a bar with typical dimensions $0.2\times
0.2\times 1cm^{3}$. Magnetoresistance measurements were performed with
magnetic fields up to 7T, the temperature ranging from 5 K to 400 K.

\section{Results}

\subsection{Neutrons}

The powder neutron diffraction pattern of $Nd_{0.5}Ca_{0.5}MnO_{3}$ ,
collected at room temperature in the paramagnetic domain, exhibits the
orthorhombic $GdFeO_{3}$ structure ($a\approx a_{p}\sqrt{2}$, $b\approx
2a_{p}$, $c\approx a_{p}\sqrt{2}$). Refinements of the room temperature
neutron diffraction patterns, carried out in the space group $Pnma$ attest
the validity of the structure resolution. The data were first analyzed with
a ' whole pattern fitting ' algorithm in order to determine accurately the
profile shape function, background and cell parameters. This preliminary
study provided a good estimate of the $R_{wp}$ and $\chi ^{2}$ that could be
reached during the structure refinement. This whole pattern fitting led to
an agreement factor $R_{wp}=4.48\%$ and $\chi ^{2}=1.42$. The refinement was
undertaken with the room temperature neutron diffraction pattern, a
temperature for which the magnetic measurements showed a paramagnetic
behavior (see the magnetic properties section). 235 Bragg peaks were used to
refine 7 positional parameters and 4 isotropic temperature factors. The
refinement converged to give an agreement factor $R_{wp}=4.84\%$ and $\chi
^{2}=1.71$. Observed, calculated and difference diffraction profiles are
shown in Figure 1. A Jahn-Teller (J-T) distortion of the $Q_{2}$ type
(antiferrodistorsive), that displaces the basal plane oxygen atoms from
their ideal positions, has been proposed to be responsible for the
insulating state. This distortion does not change the lattice symmetry ($%
Pnma $) but modifies the cell deformation in such a way that $b/\sqrt{2}\leq
c\leq a$ (the O'-type structure). Thus, the resulting Mn-O bond lengths of
1.943 \AA\ and 1.948 \AA\ (basal plane) and 1.936 \AA\ (apical distance) at
room temperature are at the origin of a small gap between the J-T split Mn $%
e_{g}$ bands ($d_{x^{2}-y^{2}}$ and $d_{z^{2}}$).

On cooling the specimen $Nd_{0.5}Ca_{0.5}MnO_{3}$ below 250 K, very weak
extra reflections appear on the electron microdiffraction patterns \cite
{Raveau}. The patterns can only be indexed by doubling the $a$ lattice
parameter. However, the extra reflections are too weak to determine the
point group from electron microdiffraction patterns and from convergent beam
electron diffraction patterns. Therefore, subsequent refinements based on
neutron diffraction data were carried out in the higher symmetry $Pnma$
space group, characteristic of the structure above the transition. It should
be clear that this procedure will yield only a structure averaged over the
symmetry operators of the $Pnma$ space group. Refinements of a $Pnma$ model
lead to large displacement parameters of the oxygen atoms indicative of
disorder and/or subtle distortions not taken into account by an orthorhombic
model. The corresponding $Nd_{0.5}Ca_{0.5}MnO_{3}$ structural parameters,
selected bond distances and angles at room temperature and 1.5 K are
reported in Table 1. The most significant difference between the room
temperature and the low temperature structures is in the Mn-O bond lengths.
At room temperature, the octahedral coordination of manganese with oxygen is
almost undistorted, with six approximately equal Mn-O distances. At low
temperature, the two $Mn-O_{1}$ distances (along the b axis) become shorter
than the four $Mn-O_{2}$ distances in the a-c plane, implying a J-T
distortion of the 'apically compressed' type as previously reported by
Radaelli et al.\cite{Radaelli} in L$a_{0.5}Ca_{0.5}MnO_{3}$. The mean
distortion in the $MnO_{6}$ octahedra $\Delta _{d}$ is increased by a factor
of 20 as the temperature decreases from room temperature to low temperature.
These results clearly demonstrate that there is a close link between the
lattice parameters and the presence of the J-T distorted $Mn^{3+}O_{6}$
octahedra with the $d_{z^{2}}$ orbital oriented in the a-c plane.

The high-flux powder diffractometer D1b was first used to investigate the
thermal dependence of the orthorhombic cell parameters in the temperature
range of 1.5 K-293 K (Figure 3). One observes that a and c parameters
increase as the temperature is lowered from $T_{CO}=250K$ to $T=160K$
accompanied by a significant decrease in the b parameter. Our results
demonstrate that the charge ordered state is progressively established from $%
T_{CO}=250K$ to $T=160K$.

The magnetic reflections in the low-temperature neutron powder diffraction
pattern have been indexed on the basis of ($a\approx a_{p}\sqrt{2}$, $%
b\approx 2a_{p}$, $c\approx a_{p}\sqrt{2}$) unit cell, previously reported
by Wollan and Koehler \cite{Wollan}. The magnetic structure of $%
Nd_{0.5}Ca_{0.5}MnO_{3}$, known as CE-type (Figure 4), is quite complex : it
entails a quadrupling of the volume of the original orthorhombic unit cell
and consists of two magnetic sublattices with independent propagation
vectors. This observation, associated with the fact that $%
Nd_{0.5}Ca_{0.5}MnO_{3}$ is an insulator (see electrical properties below),
has confirmed the hypothesis, first formulated by Goodenough \cite
{Goodenough} and developed by others \cite{Others}, that the two sublattices
result from charge ordering between $Mn^{3+}$ and $Mn^{4+}$. In Goodenough's
model, charge ordering is accompanied by orbital ordering forming
ferromagnetic zig-zag chains in the a-c plane ($d_{z^{2}}\,Mn^{3+}$ orbitals
associated with the long $Mn^{3+}-O$ bonds in the J-T distorted$Mn^{3+}O_{6}$
octahedra). These ferromagnetic chains are connected antiferromagnetically
each others in the a-c plane and along the b-axis. The magnetic structures
of the two sublattices have different propagation vectors : $\overrightarrow{%
k}_{1}=(0,0,1/2)$ for $Mn^{3+}$ and $\overrightarrow{k}_{2}=(1/2,0,1/2)$ for 
$Mn^{4+}$. Rietveld refinements of the low-temperature neutron powder
diffraction pattern clearly indicate that the magnetic moments are oriented
along the a-axis (Figure 4). The refined values of the magnetic moments at
2K are reported in Table 2 while their temperature variations are plotted in
Figure 4.\ Above $T_{N}=160K$, this magnetic moment vanishes. It is worth
noticing that the long range antiferromagnetic order is established only
well below $T_{CO}.$ It is important to note here that $T_{N}$ exactly
coincides with the temperature below which the lattice parameters become
constant (Figure 3).\ 

\subsection{Electric properties}

The electric resistivity $\rho \left( T\right) $ is plotted in Figure 6.\
The compound remains insulating what ever the temperature.\ Below 50K, the
resistivity is too high to be measured.\ Such an insulating state is in
agreement with Goodenough predictions of the CE type antiferromagnetic
state.\ The discontinuity around 250K corresponds to the charge order
transition. Magnetic fields have little effects on the resistivity, up to 7
T at least. Above $T_{CO}$, a model of small polaron describes correctly the
temperature dependence:

$\rho \left( T\right) =\rho _{0}+A\,T^{1.5\,}e^{\left( \frac{E}{T}\right) }$%
with $\rho _{0}=1.76\times \,10^{-3}$ $\Omega cm,\,A=1.40\times
10^{-8},E=947K$. At lower temperature, Below $T_{N}$, a model of variable
range hoping is more suitable: $\rho \left( T\right) =A\,e^{\left(
E/T\right) ^{0.25}}$ with $A=1.84\times 10^{-16},E=3.4\times 10^{8}K$.\ In
the intermediate temperature range, there is a gradual change from the
polaronic behavior towards the variable range hopping behavior.

\subsection{Magnetic properties}

The DC susceptibility $\chi _{DC}$ is plotted in Figure 7 as a function of
temperature. A pronounced peak is visible at 250K which corresponds to the
charge ordering temperature $T_{CO}$ established by neutron diffraction
investigation. Below $T_{CO}$, the susceptibility is reduced suggesting the
occurrence of {\bf antiferromagnetic} correlations giving no long range
order. On the other hand, no anomaly occurs at the antiferromagnetic
ordering temperature $T_{N}$ determined by neutron diffraction analysis. The
increase in $\chi _{DC}$ at low temperatures is attributed to $Nd^{3+}$ ions
and will be discussed later.

We have also undertaken high temperature susceptibility measurements up to $%
800K$ using a Faraday balance.\ The susceptibility follows a Curie-Weiss law
with a positive Curie Temperature $\theta _{1}=216K$\ \ indicative of {\bf %
ferromagnetic} interactions, (Figure 8). The Curie constant deduced from the
linear part of the $\chi _{DC}^{-1}(T)$ curve, (450 K 
%TCIMACRO{\TEXTsymbol{<} }%
%BeginExpansion
\mbox{$<$}%
%EndExpansion
T 
%TCIMACRO{\TEXTsymbol{<} }%
%BeginExpansion
\mbox{$<$}%
%EndExpansion
800 K), gives an effective moment $\mu _{eff}=$ $5.30\mu _{B}$. Assuming a
rigid coupling of the moments of the $Nd^{3+}$, $Mn^{4+}$and $Mn^{3+}$ ions,
one should measure, with 0.5 $Nd^{3+}$, 0.5 $Mn^{4+}$and 0.5 $Mn^{3+}$ions
per formula unit, $\mu _{eff}^{calc}=\sqrt{0.5.\mu
_{eff}^{2}(Nd^{3+})+0.5.\mu _{eff}^{2}(Mn^{3+})+0.5.\mu _{eff}^{2}(Mn^{4+})}$%
. The following assumptions have been made: the electronic levels of $%
Nd^{3+} $, at high temperature, are well described by $g=\frac{8}{11}$, $J=%
\frac{9}{2}$ , which leads to $\mu _{eff}=gJ\left( J+1\right) \mu
_{B}=3.62\mu _{B}$;\ for $Mn^{4+}$and $Mn^{3+}$, the orbital momentum is
quenched so that S is the appropriate quantum number: $\mu _{eff}=gS\left(
S+1\right) \mu _{B}$, with $g=2$ and $S=\frac{3}{2}$ or $2$ , which leads to 
$\mu _{eff}$ $=3.87\mu _{B}$ for $Mn^{4+}$and $\mu _{eff}$ $=4.90\mu _{B}$
for $Mn^{3+}$. This gives $\mu _{eff}^{calc}=$ $5.10\mu _{B}$, not far from
the experimental value.

DC magnetization measurements were performed up to 22 Tesla.\ Typical curves
are plotted in Figure 9, one taken above $T_{CO}$, two others between $%
T_{CO} $ and $T_{N}$ and the last one below $T_{N}$. From these data, it
seems that the compound presents a magnetic order occurring at the same time
as the charge order.\ A field induced transition, analogous to a spin flop
transition, is clearly visible below $T_{CO}$.This transition is first
order, that is, strongly hysteretic.

We have further analyzed the magnetization curves in terms of Brillouin
function, above the ''spin flop field'', by plotting the magnetization $M$
as a function of $H_{i}/T$ where $H_{i}$ is the internal field: $%
H_{i}=H_{a}+H_{m}$, $H_{a}$ is the applied field and $H_{m}$ the molecular
field due to the magnetic interactions, proportional to the magnetization.\
All the curves in the temperature range 290K-170K collapse on a single curve
at high field with the following fitting parameters: the Curie temperature
is $\theta _{2}=175K$\ and the effective moment is $5.83\mu _{B}$. It should
be noted that this analysis is made in the temperature range where $\chi
_{DC}$ slightly departs from the high temperature Curie-Weiss law, (Figure
8). This would indicate that the susceptibility could be described with two
Curie-Weiss laws, one with $\theta _{2}=175K=T_{N}$, $\mu _{eff}=5.83\mu
_{B} $ for the temperature range 170K%
%TCIMACRO{\TEXTsymbol{<}}%
%BeginExpansion
\mbox{$<$}%
%EndExpansion
T%
%TCIMACRO{\TEXTsymbol{<}}%
%BeginExpansion
\mbox{$<$}%
%EndExpansion
290K and for high magnetic fields and the other one with $\theta
_{1}=210K\simeq T_{CO},\mu _{eff}=5.30\mu _{B}$ for T%
%TCIMACRO{\TEXTsymbol{>}}%
%BeginExpansion
\mbox{$>$}%
%EndExpansion
400K.

As the temperature is lowered, the influence of the neodymium ions becomes
visible in the magnetic response of the compound: the susceptibility
increases at low temperature while the magnetization does not saturate. It
is however quite difficult to estimate exactly the neodymium contribution at
these temperature because of crystal field effects on the Nd electronic
level. For instance, in $NdGaO_{3}$\cite{NdGaO} which has a similar
perovskite structure, the J=9/2 electronic level is split into 5 Kramer
doublets separated by 137K, 133K, 363K and 183K so that only the fundamental
level with effective spin 1/2 is occupied at low temperature.\ This will
give a smaller effective moment and a strong deviation from a Curie law.
This is exactly what is observed at low temperature in the magnetic
susceptibility (see Figure 7).

\section{Discussion}

From the neutron diffraction and magnetization data, we can extract a phase
diagram (Figure 10). The paramagnetic phase (P),\ above $T_{CO}$, is a
Jahn-Teller distorted phase.\ It is an insulator with a thermally activated
behavior. At low temperature, below $T_{N}$, an antiferromagnetic phase
(AF-O) with complete charge and orbital ordering is present. Ferromagnetic
zig-zag chains of $Mn^{3+}$ and $Mn^{4+}$ are formed in the a-c plane. The
Jahn-Teller distortion of the $Mn-O_{6}$\ octahedra has considerably
increased.\ At the same time, the electrical resistivity reflects some kind
of disorder: a variable range hopping regime prevails. In the intermediate
temperature range , $Mn^{3+}$ and $Mn^{4+}$ are ordered, giving rise to a
doubling of the $a$ lattice parameter, while the orbital ordering is
progressively established. It is clear that two different charge ordered
phases are present: a charge ordered phase with no long range magnetic order
in the temperature range $T_{N}<T<T_{CO}$, and a long range
antiferromagnetic phase with complete orbital ordering for $T<T_{N}$.\ Both
phases have similar response to magnetic field: at low field, they are
characterized by a small magnetic susceptibility. At higher field, both
phases present a spin flop transition, with a first order character; the
charge ordered state is then destroyed by the magnetic field and a
ferromagnetic state prevails.

In Figure 11, we have plotted the magnetization of this ferromagnetic phase
as a function of temperature, for two different fields: $15T$ and $20T$
deduced from the $M(H)$ curves. The maximum at $T_{CO}$ has been completely
removed which confirms that it is associated with the charge ordering
transition.\ We can distinguish two different contributions to the
magnetization: one arises from the Neodymium ions , the other one from the
manganese ions.

The Neodymium ions are paramagnetic (no magnetic order has been detected by
neutron diffraction in this temperature range).\ Their magnetic moment is no
longer described by $g=\frac{8}{11}$, $J=\frac{9}{2}$ as has been already
mentioned. Using the experimental data on $NdGaO_{3}$\cite{NdGaO}, we can
estimate their contribution at 40K: $M\left( 15T\right) \simeq 0.24\mu _{B}$
and $M\left( 20T\right) \simeq 0.32\mu _{B}$ per formula unit.\
Experimentally, we get $M\left( 15T\right) \simeq 3.6\mu _{B}$ and $M\left(
20T\right) \simeq 3.8\mu _{B}$. The gap can therefore reasonably be
attributed to the Neodymium ions, which leaves a value of $3.5\mu _{B}$ for
the manganese ions : this is exactly the expected saturated magnetic moment
for $0.5$ $Mn^{4+}$and 0.5 $Mn^{3+}$.

The ferromagnetic phase has then a conventional behavior: in the
paramagnetic regime, a Curie Weiss law is observed with an effective moment $%
5.30\mu _{B}$ close to the theoretical one and with a Curie Temperature $%
\theta _{1}=216K$. The ordering temperature coincides with the Curie
temperature $T_{C}\simeq \theta _{1}$. All these results suggests an
isotropic ferromagnetic exchange interaction between each $Mn^{4+}$and $%
Mn^{3+}$ ions, in agreement with the double exchange model. At low
temperature, all the manganese magnetic moment are aligned with a net
magnetization of $3.5\mu _{B}$.\ Around the magnetic transition, the system
is best described by $\theta _{2}=175K$ and $\mu _{eff}=5.83\mu _{B}$ ,
revealing the presence of antiferromagnetic correlations.

The low field, charge ordered phase is much more unconventional : whereas
magnetization measurements suggest that $T_{N}=T_{CO}$, neutron diffraction
measurements clearly demonstrate that the antiferromagnetic order is only
established 100 K below $T_{CO}$.\ A huge peak in the susceptibility is
observed at $T_{CO}$, as well as a jump in the magnetization as a function
of field while no additional lines appear in the neutron diffraction spectra
above $T_{N}$.

We can reconciliate all the experimental results as follows.

When lowering the temperature from room temperature, the susceptibility
first increases due to the formation of a ferromagnetic phase. But this
tendency to ferromagnetic order is destroyed by charge ordering which favors
antiferromagnetism.\ The quenching of the double exchange interaction is
responsible for the susceptibility peak at $T_{CO}$ as was proposed already
in \cite{Pollert}. In the temperature range $T_{N}<T<T_{CO}$ , orbital
ordering sets in progressively; it is only achieved around 160K where long
ranger magnetic order is established. This is suggested by the temperature
variation of the lattice parameters: as the temperature is lowered from $%
T_{CO}$, they progressively decrease and remain constant only below $T_{N}$.
In this picture, when \ $T_{N}<T<T_{CO}$, the manganese ions are charge
ordered without complete orbital ordering. Several hypotheses remain.

The first one is that $T_{CO}$ corresponds to a magnetic transition towards
some unconventional magnetic order, such as modulated or helicoidal phase or
non commensurate phase, $T_{CO}\simeq T_{N1}$, and $T_{N}$ corresponds to a
commensurate magnetic order.\ There are some experimental evidences, in
electron microscopy measurements, that the charge ordered state has a non
commensurable structure in the temperature range $T_{N}<T<T_{CO}.$

The second hypothesis is that it is not a long range magnetic order but
rather fluctuations are present: ferromagnetic fluctuations above $T_{CO}$
and increasing antiferromagnetic fluctuations together with decreasing
ferromagnetic fluctuations below $T_{CO}$, giving rise to the susceptibility
peak at $T_{CO}$. A similar behavior has been observed in $%
Bi_{0.18}Ca_{0.82}MnO_{3}$\cite{Bao} : ferromagnetic fluctuations have been
observed above the charge order transition $T_{CO}$ but disappears at $%
T_{CO} $ while antiferromagnetic correlations set in. A long range
antiferromagnetic order was found at lower temperature.

A third hypothesis is that there is a mixture of the ferromagnetic phase and
the charge ordered phase. Similar mixing of phases has been observed in $%
La_{0.5}Ca_{0.5}MnO_{3}$ using NMR\cite{nmr}: antiferromagnetic and
ferromagnetic domains are found to coexist at all temperatures below the
first formation of charge ordered state.

The first hypothesis should be disregarded since no additional peak where
observed in the high resolution neutron diffraction patterns.\ Besides,
recent Electron Spin Resonance measurements on the same $%
Nd_{0.5}Ca_{0.5}MnO_{3}$ compound \cite{ESR} are in favor of the second
hypothesis: the charge ordered phase is characterized by strong magnetic
fluctuations with no long range magnetic order.

\section{Conclusions}

Combining neutron diffraction, electric transport and magnetization
measurements, we have obtained a detailed description of $%
Nd_{0.5}Ca_{0.5}MnO_{3}$ phase diagram.\ 

We have shown that, in the compound $Nd_{0.5}Ca_{0.5}MnO_{3}$ , the
paramagnetic regime above $T_{CO}=250K$ corresponds to the progressive
establishment of a ferromagnetic phase with $T_{C}\simeq \theta _{1}\simeq
T_{CO}$.

From neutron diffraction experiments, a long range antiferromagnetic phase
is observed only below $T_{N}=160K$. Then the manganese ions form
ferromagnetic zig-zag chains coupled antiferromagnetically (CE type of
ordering). The Jahn-Teller distortion is greatly enhanced and complete
orbital ordering is established. The coumpound remains insulating.

A more complicated phase exists between $T_{N}$ and $T_{CO}$. In this phase,
orbital ordering sets in progressively.\ antiferromagnetic and ferromagnetic
interactions compete. At low magnetic field, a non magnetic, insulating
state prevails whereas at high magnetic field the ferromagnetic state is
favored. The transition towards the ferromagnetic phase, from the
non-magnetic as well as the magnetic charge ordered phase is strongly
hysteretic revealing a first order type transition.

\section{Acknowledgment}

One of the authors (F.Millange) thanks O.\ Isnard and E. Suard for the
neutron diffraction data collected on D1B diffractometer. We aknowledge also
V.\ Caignaert for helpful discussion on the neutron diffraction results. The
Grenoble High Magnetic Field Laboratory is 'laboratoire conventionn\'{e}
\`{a} l'Universit\'{e} Joseph Fourier'.

\section{Figure and table captions}

{\bf Table 1}: $Nd_{0.5}Ca_{0.5}MnO_{3}$ structural parameters at room
temperature (RT) and 1.5K, as determined from Rietveld refinements based on
neutron powder diffraction data.

{\bf Table 2}: Refined magnetic moment components of the $Mn^{3+}$ and $%
Mn^{4+}$ sublattices. the site coordinates are expressed as fractions of the
($2a\times b\times 2c$) magnetic cell.

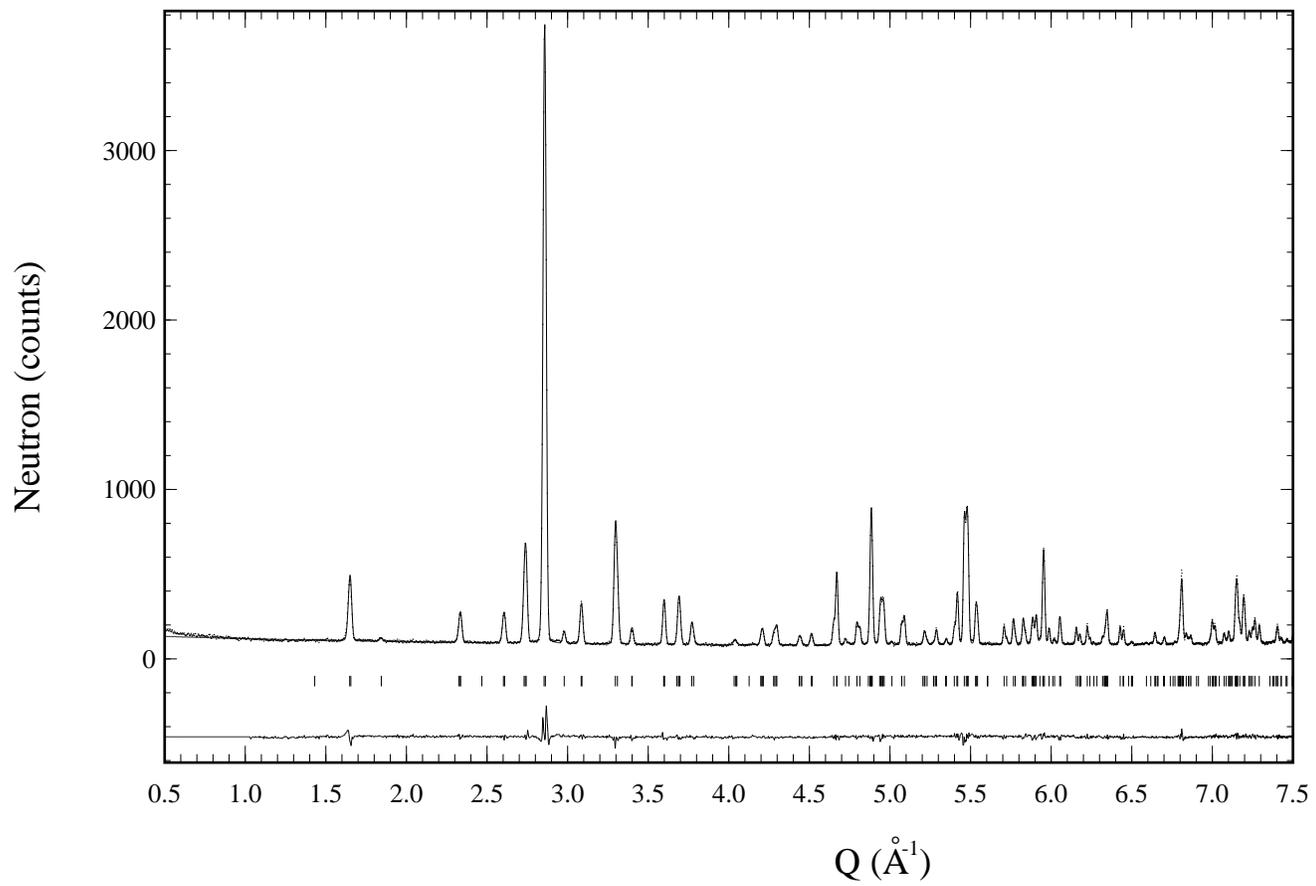
{\bf Figure 1}: Rietveld refinement plot of the neutron diffraction data at
room temperature. The ticks are given for the nuclear structure peaks.

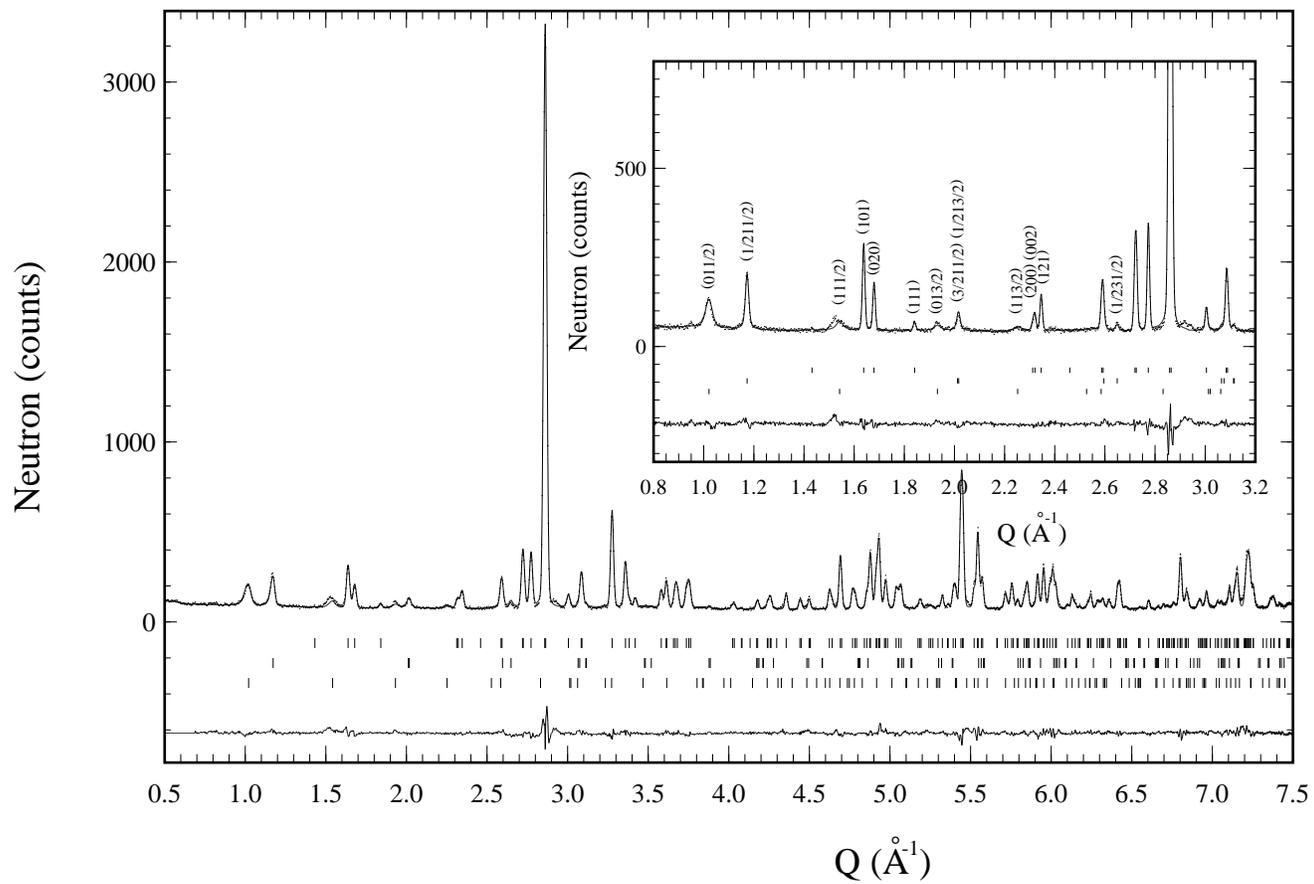
{\bf Figure 2:} Rietveld refinement plot of the neutron diffraction data at
1.5 K. The ticks are given for the nuclear structure peaks (upper), magnetic
structure peaks ( $Mn^{4+}$sublattice, middle) and magnetic structure peaks
( $Mn^{3+}$ sublattice, lower). Points are the experimental data, line is
the Rietveld fit and the lower line is the difference curve.

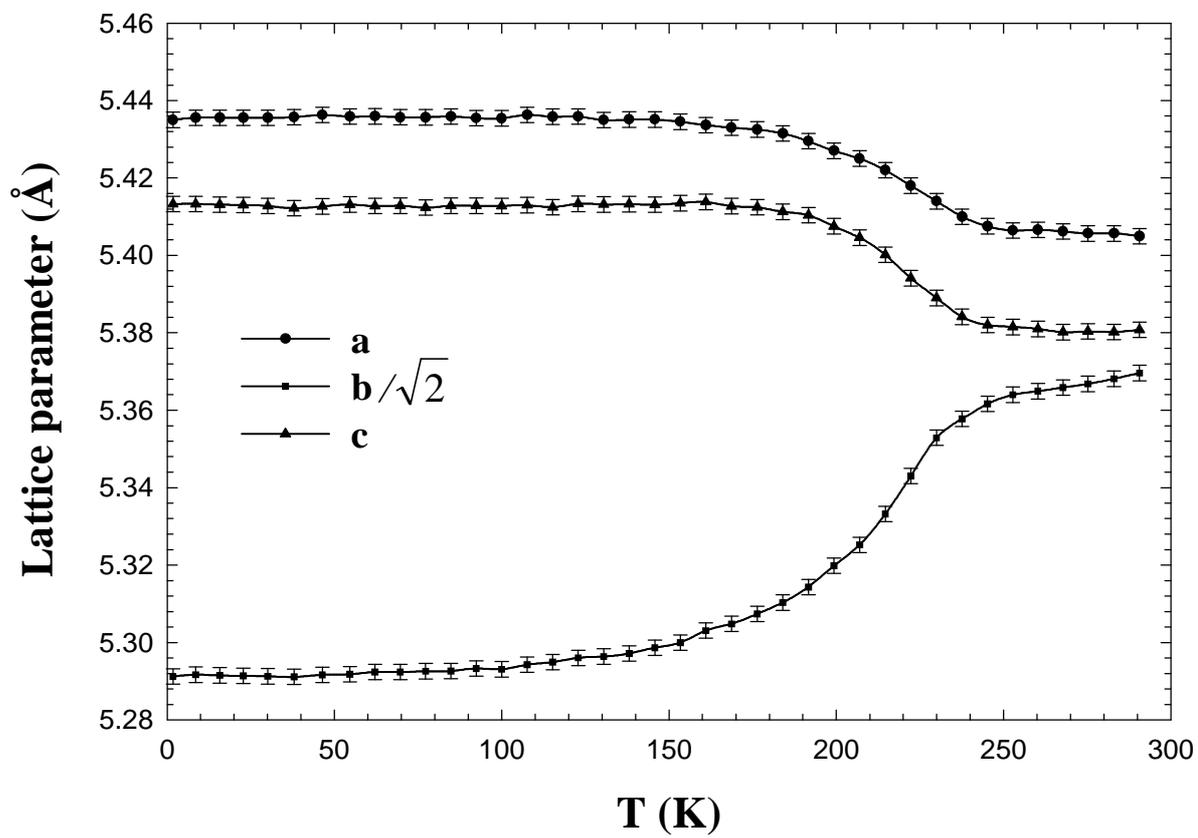
{\bf Figure 3}: Lattices parameters of $Nd_{0.5}Ca_{0.5}MnO_{3}$ as a
function of temperature. The charge order transition $T_{CO}$ is indicated
by an arrow.

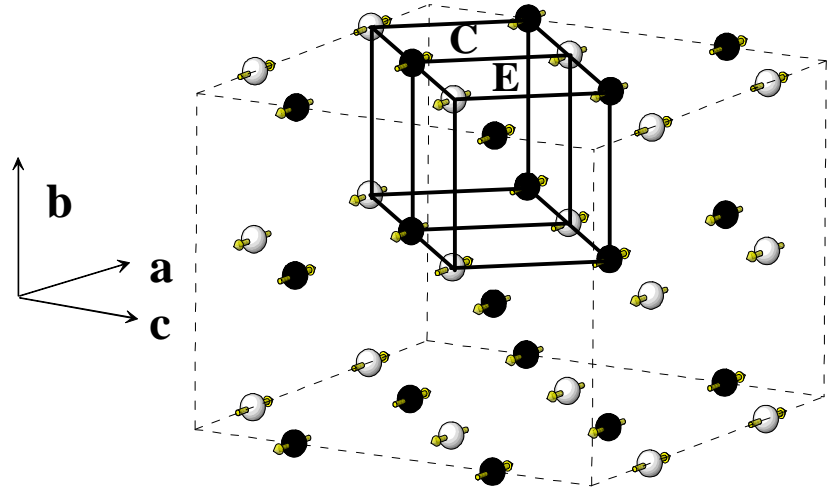
{\bf Figure 4}: Representation of the CE-type magnetic structure. This
structure is characterized by two Mn sublattices forming ferromagnetic
zig-zag chains in the a-c plane.

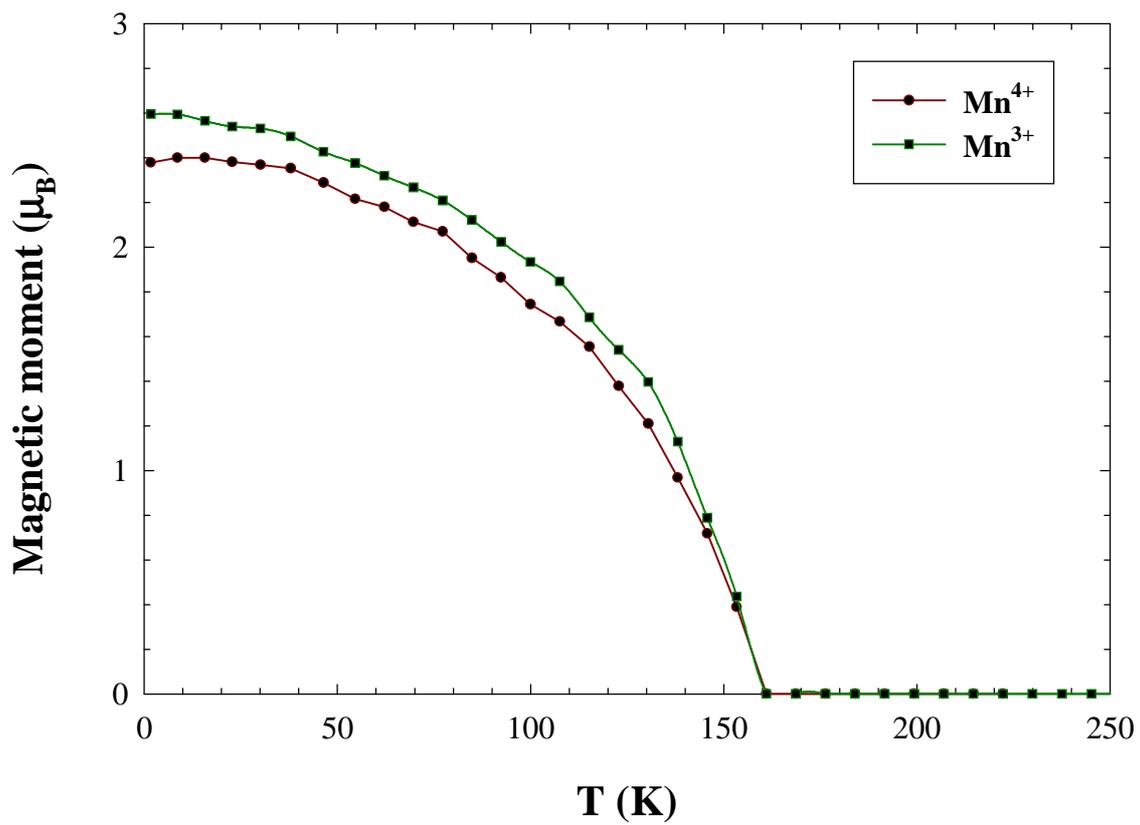
{\bf Figure 5}: Magnetic moment of the manganese ions as a function of
temperature.

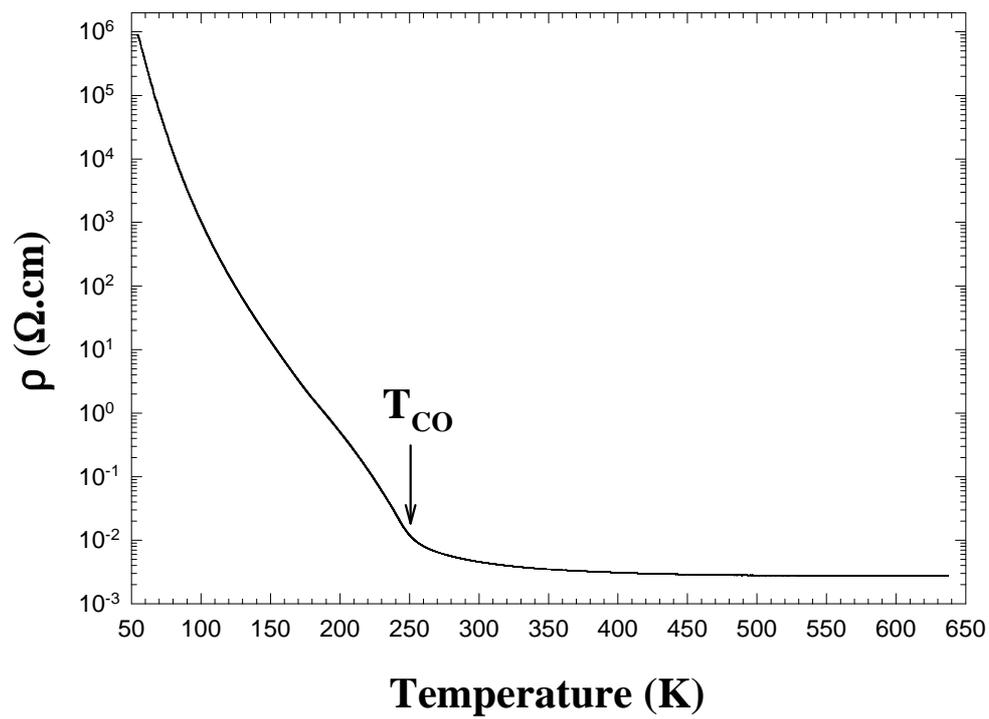
{\bf Figure 6}: Electric resistivity as a function of temperature. The
antiferromagnetic transition $T_{N}$ is indicated by an arrow.

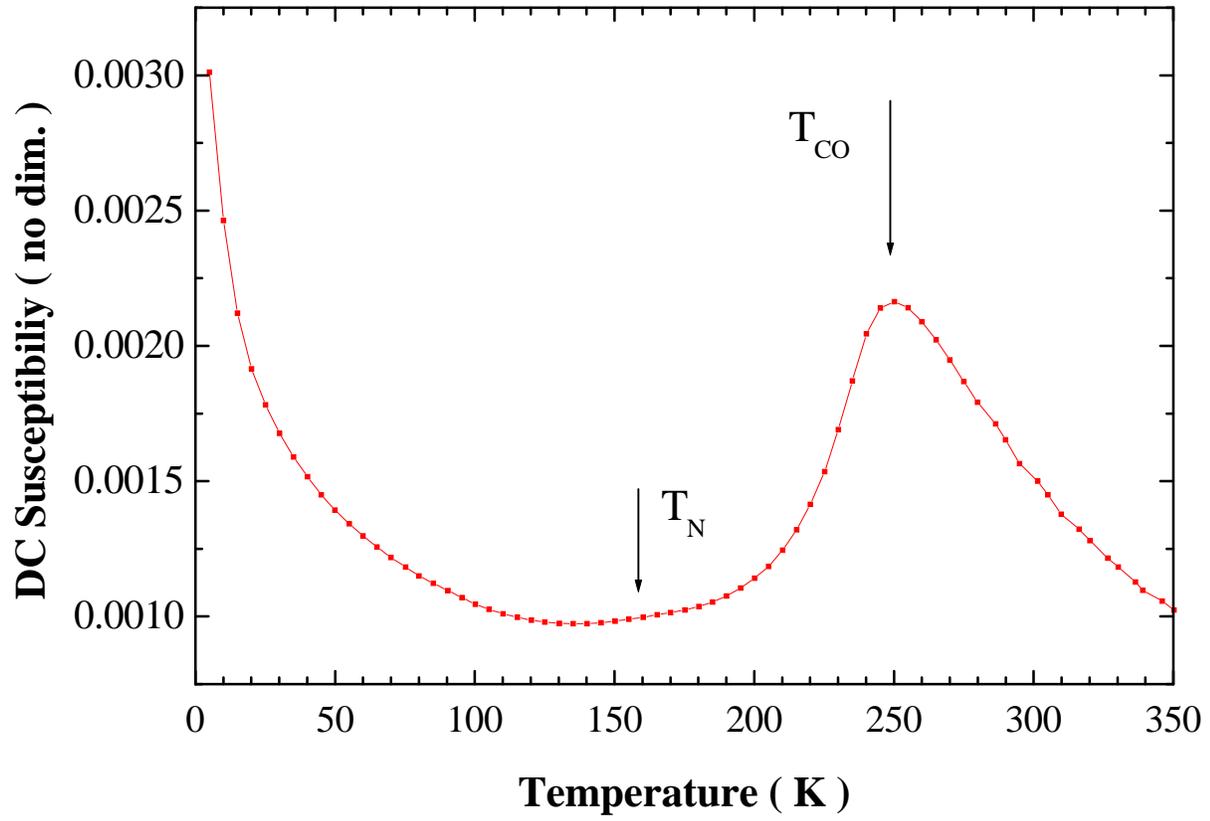
{\bf Figure 7}: DC susceptibility (10mT) as a function of temperature.

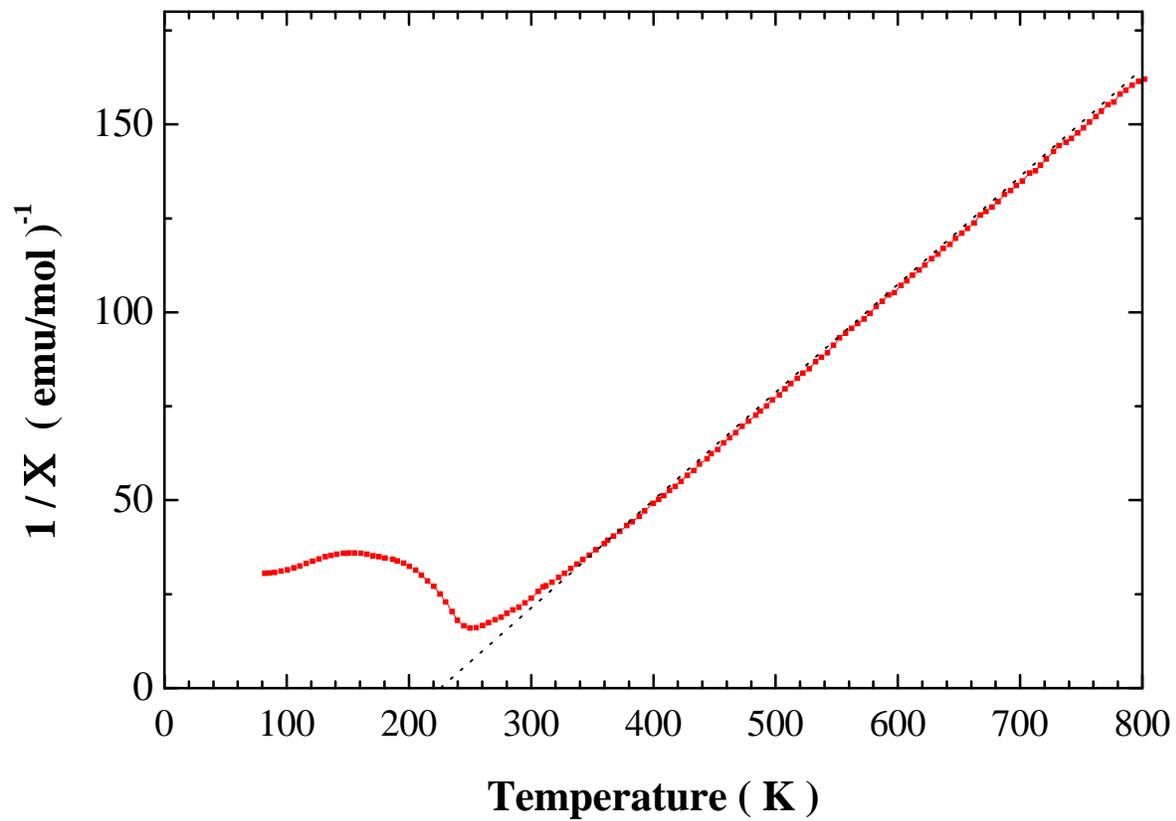
{\bf Figure 8}: Inverse of the susceptibility at high temperature. The
dotted line is a linear fit giving a Curie temperature $\theta _{1}=216K$
and an effective moment $\mu _{eff}=$ $5.30\mu _{B}$.

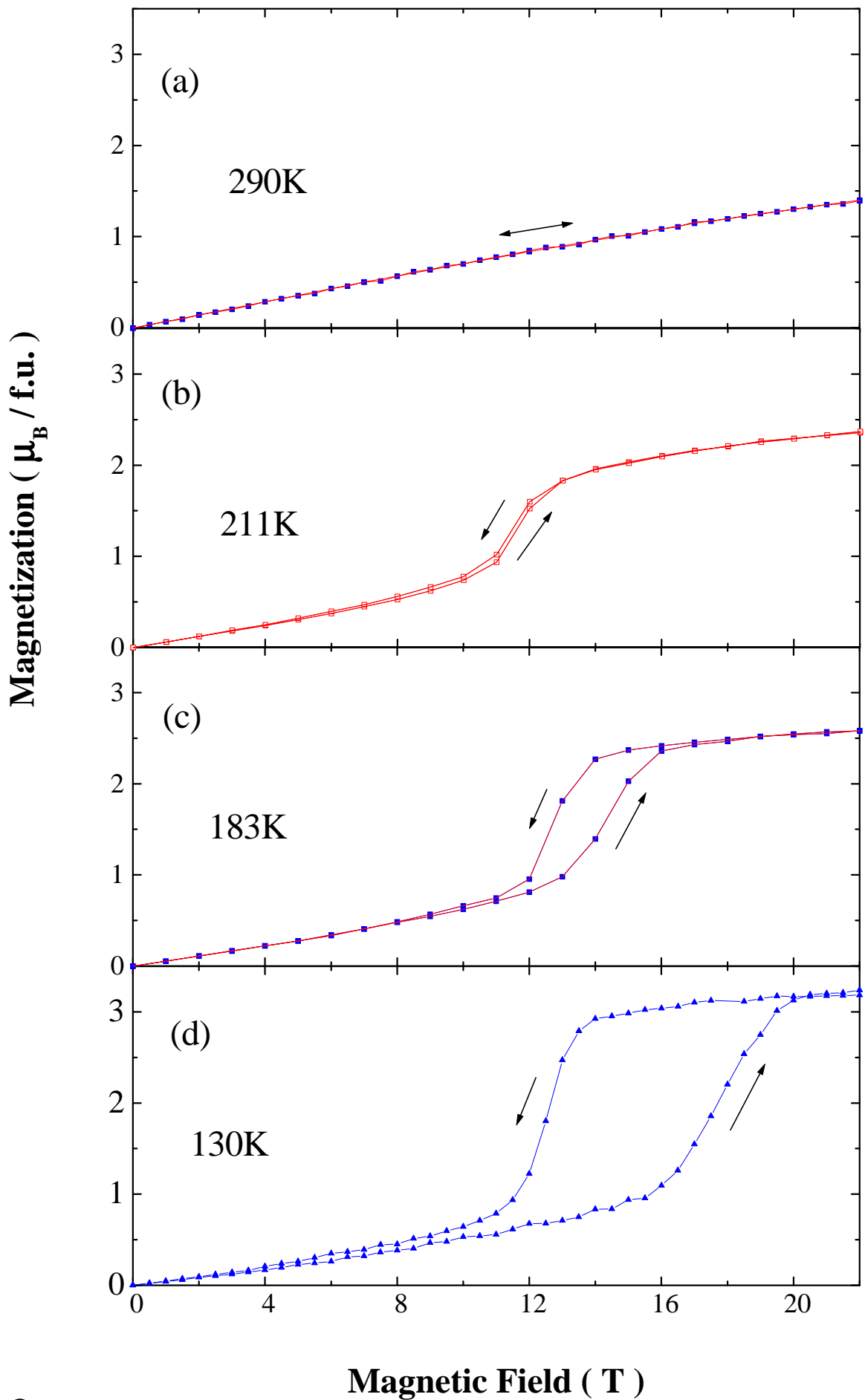
{\bf Figure 9}: DC magnetization as a function of magnetic field:

(a) $T=290K$ (paramagnetic domain)

(b) $T=211K$ and (c) $T=183K$ (charge order domain)

(d) $T=130K$ (antiferromagnetic domain)

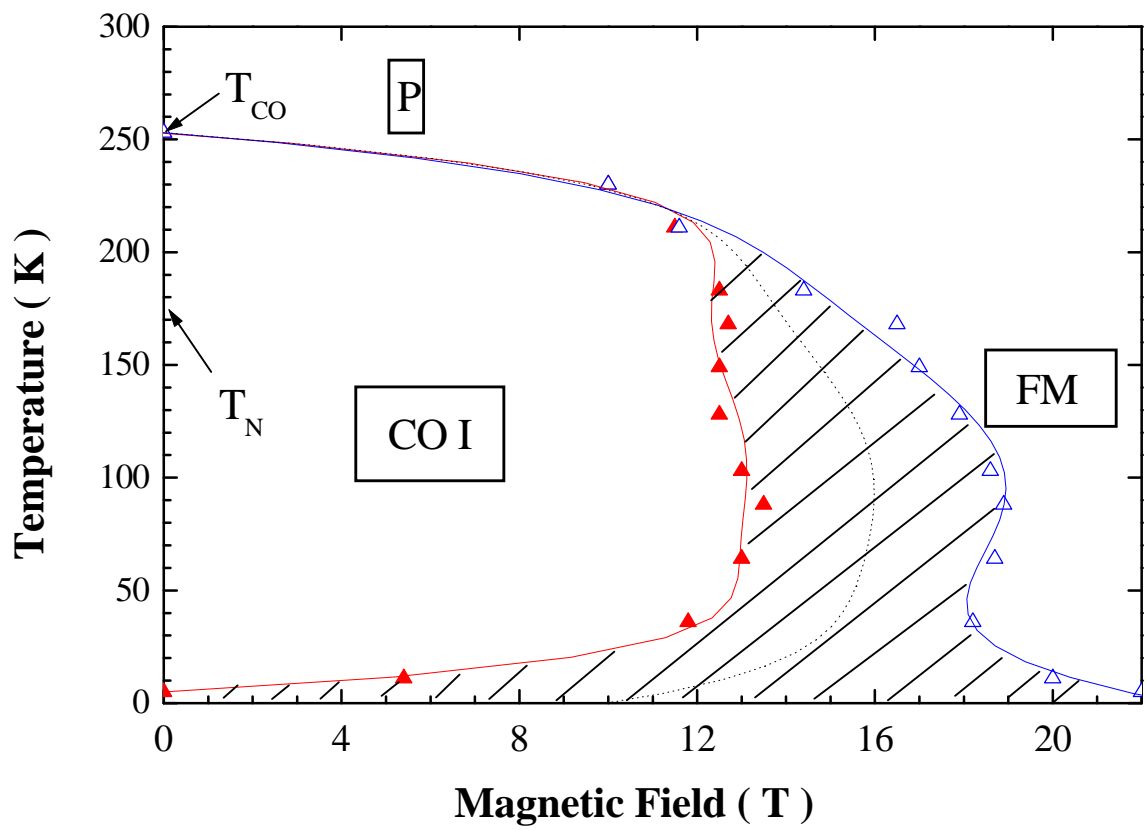
{\bf Figure 10}: Magnetic phase diagram of $Nd_{0.5}Ca_{0.5}MnO_{3}$. $%
T_{CO} $ and $T_{N}$ are determined from neutron diffraction data. The
different lines are guides to the eyes to separate the different phases :
paramagnetic (P), charge ordered (CO), antiferromagnetic, orbital ordered
(AF-O) and ferromagnetic (FM) phases. The dotted line corresponds to the
average spin flop field.

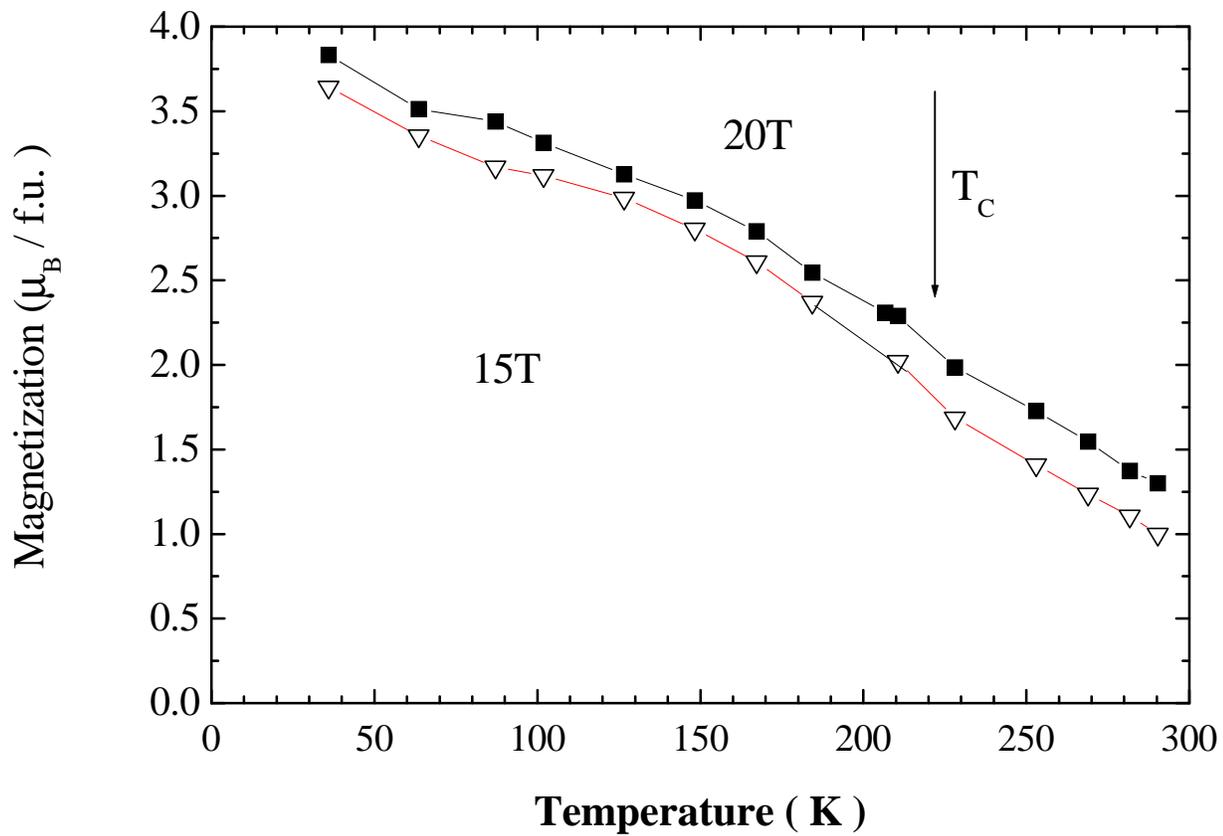
{\bf Figure 11}: Magnetization as a function of temperature for the high
field ferromagnetic phase at 20T and 15T.

\end{document}